\documentclass[12pt]{article}

\usepackage{scicite}
\usepackage{authblk}
\usepackage{times}
\usepackage{amsmath} 
\usepackage{graphicx}

\newenvironment{sciabstract}{%
\begin{quote} \bf}
{\end{quote}}

\title{Higher-order topology in monolayer graphene}

\author[1,2]{Feng Liu\thanks{LiuFeng@nbu.edu.cn}}
\author[3,4]{Katsunori Wakabayashi\thanks{waka@kwansei.ac.jp}}
\affil[1]{School of Physical Science and Technology, Ningbo
  University, Ningbo, 315-211, China}
\affil[2]{Laboratory of Clean Energy Storage and Conversion, Ningbo
  University, Ningbo, 315-211, China}
\affil[3]{Department of Nanotechnology for Sustainable Energy, School of Science and Technology, Kwansei Gakuin University, Gakuen 2-1, Sanda 669-1337, Japan}
\affil[4]{Center for Spintronics Research Network (CSRN), Osaka University, Toyonaka 560-8531, Japan}

\date{}

\begin{document} 


\baselineskip24pt


\maketitle


\begin{sciabstract}
  We show that monolayer graphene intrinsically hosts higher-order
  topological corner states, in which electrons are localized
  topologically at atomic sizes.
  The emergence of the topological corner states in graphene is due to
  a nontrivial product of the Zak phases for two independent
  directions, which can be handily calculated graphically by using the bulk
  wavefunctions. We give an explicit expression that indicates the
  existence of topological corner states for various geometric edges
  and corner angles.
  We also demonstrate the nontrivial localization nature of the
  topological corner states in graphene by putting an imaginary onsite
  potential mask.   
\end{sciabstract}

\clearpage
The topology of energy bands has offered us a new dimension of designing
solid-state materials with intriguing properties~\cite{Bansil2016}.
One of the essential properties of topologically nontrivial systems is the bulk-edge
correspondence, where robust edge and surface states appear at the
interfaces which separate two topologically distinct systems~\cite{Hatsugai1993,Haldane2006}.
These topological states are insensitive to the local perturbations that
preserve the bulk topological invariants, also called topological protection. 
The recently proposed higher-order topology has extended such the
bulk-edge correspondence to the more general bulk-edge-corner
correspondence,
where topological states of co-dimension larger than 1, i.e.,
topological corner states in two-dimensional systems, appear~\cite{Benalcazar2017,Song2017}. 
The higher-order topology has attracted significant attention in
condensed matter physics, not only for its fundamental scientific
importance but also because of its potential applications in
electronics.
In particular, topological corner states provide us the possibility of designing laser
cavity and quantum computation with topological protection and maximized efficiency~\cite{Hararieaar2018,Zhang2020,Wu2020B,Wu2020}.

On the basis of several prototype models of higher-order topology,
such as the two-dimensional Su-Schrieffer–Heeger model, the quadrupole insulator and the breathing Kagome lattice,
nontrivial higher-order topology usually requires the fine-tuning of hopping
textures plus external fields~\cite{Liu2017,Benalcazar2017B,Ezawa2018}.
Along this line, several proposals, such as monolayer
graphdiyne~\cite{LiuBing2019}, Kekul\'e-like lattice~\cite{Liu2019},
twisted bilayer graphene~\cite{Park2019}, and topological insulators with
the breaking of time-reversal symmetry~\cite{Ren2020}, have been made.
Unfortunately, the fine-tuning of hopping textures is difficult to
realize in solid-state materials, as difficulties in the precise control
of crystal growth.
Besides the observation of the topological corner states in several artificial
crystalline
structures~\cite{Serra-Garcia2018,Peterson2018,Imhof2018,Ota2019,Xue2020,Xie2020},
the realization of higher-order topological states is remaining elusive in solid-state materials, especially in two-dimensions.

Without the fine-tuning of hopping textures or applying external
fields, it is hard to imagine the emergence of higher-order
topological states, especially in uniform systems.
Motivated by a simple picture that corners are ``edges'' of edges, we consider
monolayer graphene as a possible candidate for topological corner
states. Because in graphene, edge states accompanied with the perfectly
conducting channel appear for various geometric
ribbons only except for the armchair edge~\cite{Waka2009, Akhmerov2008}. 
Later it was shown that the emergence of
these edge states in graphene is due to a nonzero geometric phase -- Zak phase~\cite{Delplace2011}.
The Zak phase is a topological indicator for systems of zero Chern
number that corresponds to the bulk charge
polarization~\cite{Zak1989,Vanderbilt1993,Obana2019,Song2020}.
Finite bulk charge polarization casts fractional surface charge on the
direction perpendicular to the edges and results topological edge states~\cite{Vanderbilt1993}.  
When the Zak phases in graphene along the two directions those are
perpendicular to the edges forming the
corner are both nontrivial, a corner state emerges.

We employ the tight-binding model with nearest-neighbor hoppings to
describe the $\pi$-electronic states of graphene. The Hamiltonian for bulk graphene can be
written as 
\begin{equation}
    H(\mathbf{k})=
   -t|\rho(\mathbf{k})|
  \begin{pmatrix}
    0 & e^{-i\phi(\mathbf{k})}\\
    e^{i\phi(\mathbf{k})} & 0
  \end{pmatrix}
\end{equation}
in the basis of the two sublattices A and B as shown in Fig.~1A. Here, 
$\rho(\mathbf{k})=1+e^{-i\mathbf{k}\cdot \mathbf{a}_1}+e^{-i\mathbf{k}\cdot
  \mathbf{a}_2}\equiv |\rho(\mathbf{k})|e^{-i\phi(\mathbf{k})}$.  
$\mathbf{a}_1=a(1/2,\sqrt{3}/2)$ and $\mathbf{a}_2=a(-1/2,\sqrt{3}/2)$ are
the primitive vectors, where $a=2.46$ \AA\ is lattice constant of graphene.
$\mathbf{k}=(k_x,k_y)$ is the wavenumber vector. 
$t=2.75$ eV is the electron transfer integral between nearest-neighbor carbon
atoms. The eigenenergies of bulk graphene are composed of two
bands
$E_\pm=\pm t|\rho(\mathbf{k})|$, and the corresponding eigen-wavefunctions
are $|u_\mathbf{k},\pm\rangle=\frac{1}{\sqrt{2}}(e^{-i\phi(\mathbf{k})},\pm 1)^T$.
For the valence band states $|u_\mathbf{k},-\rangle$, the Zak
phase along a specific direction $\mathbf{i}$ is given as
\begin{equation}
  Z_\mathbf{i}(k_\mathbf{j})=\dfrac{1}{2}
  \int_{\mathbf{P}_\mathbf{i}(k_\mathbf{j})} \dfrac{d \phi (\mathbf{k})}{d k_{\mathbf{i}}} d k_{\mathbf{i}},
\end{equation}
where the direction $\mathbf{j}$ is perpendicular to $\mathbf{i}$, and
$\mathbf{P}_\mathbf{i}(k_\mathbf{j})$ is a straight path connecting two equivalent $k_\mathbf{j}$
points along $\mathbf{i}$. Because of
the chiral symmetry, the value of Eq.~(1) is either $\pi$ or $0$ in graphene. From
Eq.~(1) we see that $Z_\mathbf{i}(k_\mathbf{j})$ is determined by the winding number
of $\phi(\mathbf{k})$ along $\mathbf{P}_{\mathbf{i}}(k_\mathbf{j})$, which allows us to calculate Eq.~(1)
graphically through the density plot of $\phi(\mathbf{k})$ as displayed in Fig.~1B. For a given $k_\mathbf{j}\in[-\pi/|\mathbf{T}|,\pi/|\mathbf{T}|]$
with that $\mathbf{T}$ is the period of the graphene ribbon,
$Z_\mathbf{i}(k_\mathbf{j})=\pi$ if $\mathbf{P}_\mathbf{i}(k_\mathbf{j})$ passes through the
discontinuities of $\phi(\mathbf{k})$, and $Z_\mathbf{i}(k_\mathbf{j})=0$ otherwise.
For the typical zigzag ribbon, we have
$\mathbf{T}=\mathbf{a}_2$ and $\mathbf{P}(0)=\mathbf{b}_1$ with
$\mathbf{b}_i\cdot \mathbf{a}_j=2\pi\delta_{ij}$, as shown in Fig.~1B. For
other geometric ribbons, we have that 
$\mathbf{T}\times\mathbf{P}=\mathbf{b_1}\times\mathbf{b}_2$ in general.
By fixing the gauge, let $\mathbf{T}$ be evenly split by the
$\Gamma$ point, as in Fig.~1B.
Through Fig.~1B we see that the Zak phase is $\pi$ for
$\frac{2\pi}{3}<|k_\mathbf{j}|\leq \pi$ of the zigzag ribbon, consistent with the
energy band structure of the zigzag ribbon and the numerical calculation
of the Zak phase along $\mathbf{b}_1$ direction as displayed in Fig.~1C.
The inset of Fig.~1C displays a zigzag edge with
$\mathbf{T}=\mathbf{a}_2$. Nonzero $Z_\mathbf{i}(k_\mathbf{j})$ yields an imaginary
solution $k_\mathbf{i}=\pi+ i \eta_\mathbf{i}$ at a specific
$k^\prime_\mathbf{j}$s range for the boundary
condition,
e.g., $\sin(k_\mathbf{i}N)+2\cos(k_\mathbf{j}/2)\sin[k_\mathbf{i}(N+1)]=0$
for a width-$N$ zigzag ribbon, where $\eta_\mathbf{i}$ is the inverse of the localization length of
the edge state~\cite{Delplace2011,Wakabayashi2010}.
Equation (2) describes the first-order topology of graphene.

For the second-order topology of graphene,
we introduce the Zak phase $Z_\mathbf{m}(k_\mathbf{n})$ along another direction $\mathbf{m}$,
where the two edges those are periodic along $\mathbf{n}$ and $\mathbf{j}$ forming the corner with an angle $\theta$.
The emergence of a topological corner state suggests an
exponentially localized solution along $\mathbf{i}$ and
$\mathbf{m}$, which requires that both $Z_\mathbf{i}(k_\mathbf{j})$ and
$Z_\mathbf{m}(k_\mathbf{n})$ are nontrivial. This yields a twisted
topological structure in momentum space associated with the corner
angle $\theta$.
Without losing generalities, we set $Z_\mathbf{i}$ as nontrivial, and we have
$k_{\mathbf{i}}=\pi+i\eta_\mathbf{i}$ for the corresponding $k^\prime_\mathbf{j}$s.
Then according to the corner angle, we have $k_\mathbf{n}=\pi\sin\theta+k_\mathbf{j}^\prime\cos\theta$.
If $k_\mathbf{n} \in k_\mathbf{n}^\prime$s those give nonzero $Z_\mathbf{m}$, the
topological corner state appears in graphene.
In a compact form, such the twisted higher-order topological structure in
momentum space can be presented as
\begin{equation}
  Z^{\theta}_\mathbf{im}=
  \underset{k_\mathbf{j}>0}
  {\vee}\dfrac{Z_\mathbf{i}(k_\mathbf{j})Z_\mathbf{m}(k_\mathbf{j}\cos\theta+\pi\sin\theta)}{\pi^2},  
\end{equation}
where $\underset{k_\mathbf{j}>0}{\vee}$ indicates taking the logical ``or'' for all
possible $k_j>0$ within the 1st Brillouin zone.
Equation (3) involves two directions of the Zak phases with a twisted
angle $\theta$ that describes the second-order topology of graphene.
When $Z^{\theta}_\mathbf{im}$ is not zero, a solution of
$\mathbf{k}=(\pi+i\eta_\mathbf{i})\mathbf{i}+(\pi+i\eta_\mathbf{j})\mathbf{j}$ exists.

For the demonstration of Eq.~(3), we take the zigzag-zigzag corner as an
example. There are two possible angles for the zigzag-zigzag corners,
which are $\pi/3$ and $\pi/6$~\cite{Yuji2011}. Topological corner states appear
in the $\pi/6$ zigzag-zigzag corners and are absent in the $\pi/3$ corners, as displayed in Fig.~2A. The absence
of the corner state in the $\pi/3$ zigzag-zigzag corner is because for
the $\pi/3$ zigzag-zigzag corner, with a nontrivial Zak phase along
$\mathbf{i}$ the possible range of $k_\mathbf{n}$ is determined by the
corner angle $\pi/3$ and $k^\prime_\mathbf{j} \in (2\pi/3,\pi] $ as
\begin{equation}
  \begin{split}
    k_\mathbf{n} & \in
    [\pi\cos\frac{\pi}{3}+\pi\sin\frac{\pi}{3},\frac{2\pi}{3}\cos\frac{\pi}{3}+\pi\sin\frac{\pi}{3}) \\
    & \approx [0.37\pi,0.53\pi),
    \end{split}
  \end{equation}
which has no overlap with $k_\mathbf{n}^\prime$ range $[-\pi,-2\pi/3)\cup(2\pi/3,\pi]$ for the nontrivial Zak phase along $\mathbf{n}$. This gives a trivial $Z^\theta_\mathbf{im}$ and suggests that it is impossible to take imaginary solutions of $k_\mathbf{m}$ and $k_\mathbf{i}$ simultaneously, and thus the corner state is absent in
the $\pi/3$ zigzag-zigzag corner. We replace one zigzag edge by the
bearded edge in Fig.~2A, which has a complementary
$k^\prime_\mathbf{n}$ range $[-2\pi/3,2\pi/3]$ to the zigzag
ribbon, and we see that the topological corner states appear in the $\pi/3$
zigzag-bearded corners as displayed in Fig.~2A.

Because of the chiral symmetry, the corner state in graphene is fixed at zero
energy buried in the bulk and edge states, also called
a bound state in continuum~\cite{Benalcazar2020}. 
To pick those corner states out,
one possible way is to use a non-hermitian mask, i.e.,
an imaginary onsite potential $\alpha=-0.005t$ in the diagonal terms of
Eq.~(1), indicted by the red shadow in Fig.~2A.
With such the non-hermitian mask, only topological corner states
remain non-decayed while the bulk and edge states fade out, because of the
localization nature of the corner states.
As displayed in Fig.~2B, the decaying rate of the
topological corner state in the $\pi/6$ zigzag-zigzag corner exponentially approaches to zero 
by increasing the area of the non-hermitian mask. From the complex
energy spectrum of the $\pi/6$ zigzag-zigzag corner sample~(upper panel of Fig.~2C), we see that except the topological
corner states, other eigen-states all have finite decaying rates.
In the $\pi/3$ zigzag-zigzag corner sample,
there is no non-decayed state due to the absence of topological corner states ~(lower panel of Fig.~2C).
It is noted that except the armchair graphene ribbon, the Zak phase is always nonzero
for a range of $k^\prime_j$, and thus topological corner states exist for various
edge shapes and corner angles in general.

At last, we emphasize that the emergence of topological corner
states in graphene is due to the nontrivial Zak phases along two
non-parallelled directions. The perturbations
those break the chiral symmetry such as the unsymmetric onsite potential
and the real next-nearest-neighbor hopping would shift the eigenenergy of the
corner state, and cannot suppress their existence as far as Eq.~(3) remains nontrivial~\cite{Peng2020}.
As displayed in Fig.~3A and upper panel of Fig.~3B, with finite
unsymmetric onsite potential,
the topological corner state in the $\pi/6$ zigzag-zigzag corner
survives with a nonzero eigenenergy.
For the finite real next-nearest-neighbor hopping, it renders the corner
states with a longer decaying distance.
Figure 3C displays the density of wavefunction of the corner 
states with next-nearest hopping $0.1t$.
By increasing the size of the non-hermitian mask, we can confirm the
localization nature of the corner state in Fig.~3C.
As displayed in Fig.~3D, the imaginary part of eigenenergy of
the corner state in Fig.~3C approaches exponentially to 0 with the
size of the non-hermitian mask increasing.
For the large spin-orbit coupling, it
destroys the topological corner states in graphene, because a
finite Chern number for a single spin channel breaks the monodromy of the Zak phase.
For a single spin channel of graphene, the spin-orbit coupling can be
simulated by the Haldane model~\cite{Haldane1988,Kane2005},
where a finite imaginary next-nearest-neighbor hopping is assumed.
As shown in lower panel of Fig.~3B, the complex energy spectrum of the $\pi/6$
zigzag-zigzag corner sample with the non-hermitian mask and finite imaginary
next-nearest-neighbor hopping, the non-decaying corner states are
suppressed and a plateau of chiral edge states appear instead.
It is fortunate that the spin-orbit couplings are minimal in graphene, and we can neglect them safely.

To summarize, we have studied the higher-order topology of
monolayer graphene. We have found that topological corner states exist in
graphene for various geometric boundaries and corner angles. These
topological corner states in graphene correspond to a twisted higher-order
topological structure associated with the Zak phases in momentum space. Our results
provide a possible way of localizing electrons at atomic sizes
intrinsically in semi-metallic systems. Our predication is only based
on monolayer graphene without substrate effects and the spin-orbit
couplings, we believe that the experimental observation of topological
corner states in graphene is possible.

\section*{Acknowledgments}
F.L. acknowledges the financial support by the Research Starting Funding of Ningbo University. K.W. acknowledges the financial support by JSPS KAKENHI Grant No. JP18H01154, and JST CREST Grant No. JPMJCR19T1.

\clearpage
\bibliography{references}
\bibliographystyle{science}
\clearpage

\begin{figure}[!htp]
\begin{center}
\leavevmode
\includegraphics[clip=true,width=1.0\columnwidth]{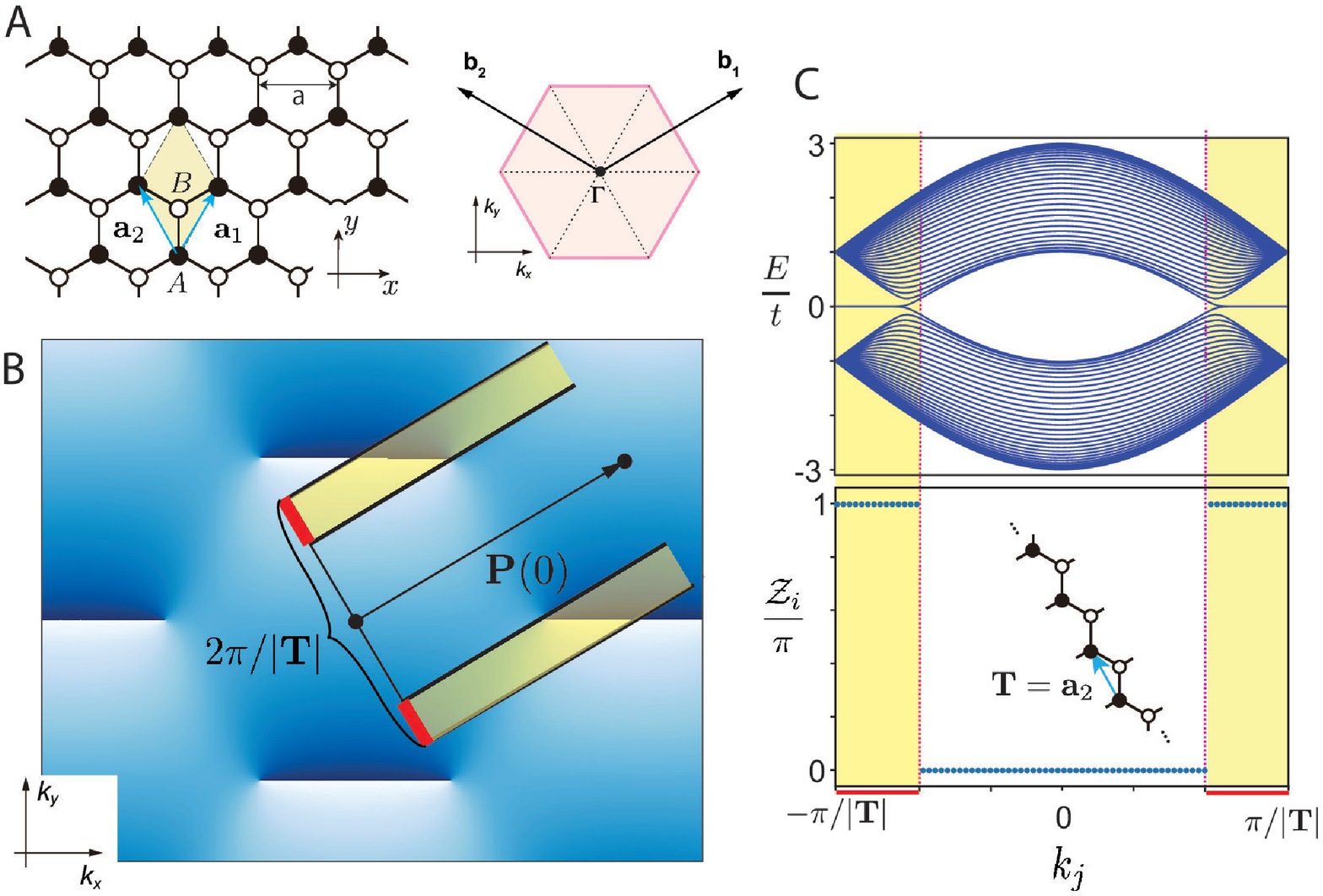}
\caption{(A) (Left) Schematic of monolayer graphene with the primitive
  lattice vectors $\mathbf{a}_1=a(1/2,\sqrt{3}/2)$ and
  $\mathbf{a_2}=a(-1/2,\sqrt{3}/2)$. $a$ is the lattice constant of graphene.
  The yellow shade of rectangle
  indicates the unit cell. 
  (Right) 1st Brillouin zone with the reciprocal primitive vectors 
  $\mathbf{b}_1=\frac{2\pi}{a}(1,1/\sqrt{3})$
  and $\mathbf{b}_2=\frac{2\pi}{a}(-1,1/\sqrt{3})$. 
  (B) Density plot of $\phi(\mathbf{k})$ in the momentum space, where the black circles indicate the $\Gamma$
  point. For a graphene zigzag ribbon with the period
  $\mathbf{T}=\mathbf{a}_2$, $\mathcal{Z}(k)=\pi$ when the path
  $\mathbf{P}(k)$ passes through the discontinuities of
  $\phi(\mathbf{k})$, as indicated by the red shades.
  (C) Energy band structure and numerical calculation of Zak phase for the zigzag graphene
  ribbon, where the nontrivial range (indicated by yellow shadow) $k^\prime_j$ is
  $[-\pi,-2\pi/3)\cup(2\pi/3,\pi]$. The width of zigzag ribbon is set as
 $15\sqrt{3}a$. 
  Inset of (C) is the schematic of the graphene zigzag edge with the
  chosen unit cell in (A) and the period $\mathbf{T}=\mathbf{a}_2$.}
\end{center}
\end{figure}

\begin{figure}[!htp]
\begin{center}
\leavevmode
\includegraphics[clip=true,width=1.0\columnwidth]{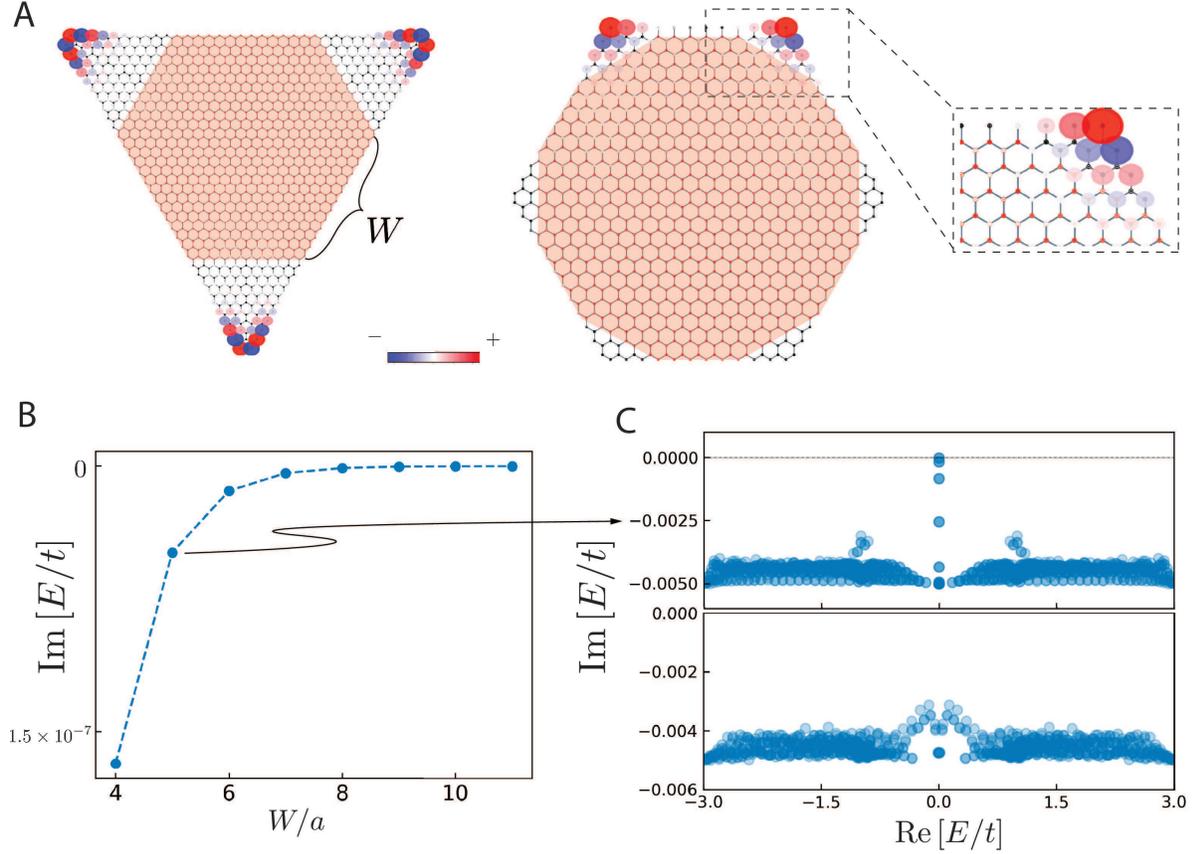}
\caption{(A) Topological corner states of the $\pi/6$ zigzag-zigzag 
  corner and the $\pi/3$ zigzag-bearded corner, where the size of
  circles indicates the local amplitude of the wavefunction, and blue
  and red indicate their signs. The red shade and sites in the middle
  parts indicate a non-hermitian mask where an imaginary onsite potential
  $-0.005t$ is assumed.
  Inset shows a zoom of the topological corner state in the  $\pi/3$
  zigzag-bearded corner.
  (B) Dependence of imaginary part of the eigenenergy for the topological
  corner state in the $\pi/6$ zigzag-zigzag corner on the
  non-hermitian mask size $W/a$.
  By increasing the size of the non-hermitian
  mask, the imaginary part of the eigenenergy of the topological corner state
  in the $\pi/6$ zigzag-zigzag corner approaches to zero exponentially.
  (C) Complex energy spectra of the $\pi/6$ zigzag-zigzag corner
  sample for the non-hermitian mask size $W/a=5$ and the $\pi/3$
  zigzag-zigzag corner sample.}
\end{center}
\end{figure}

\begin{figure}[!htp]
\begin{center}
\leavevmode
\includegraphics[clip=true,width=1.0\columnwidth]{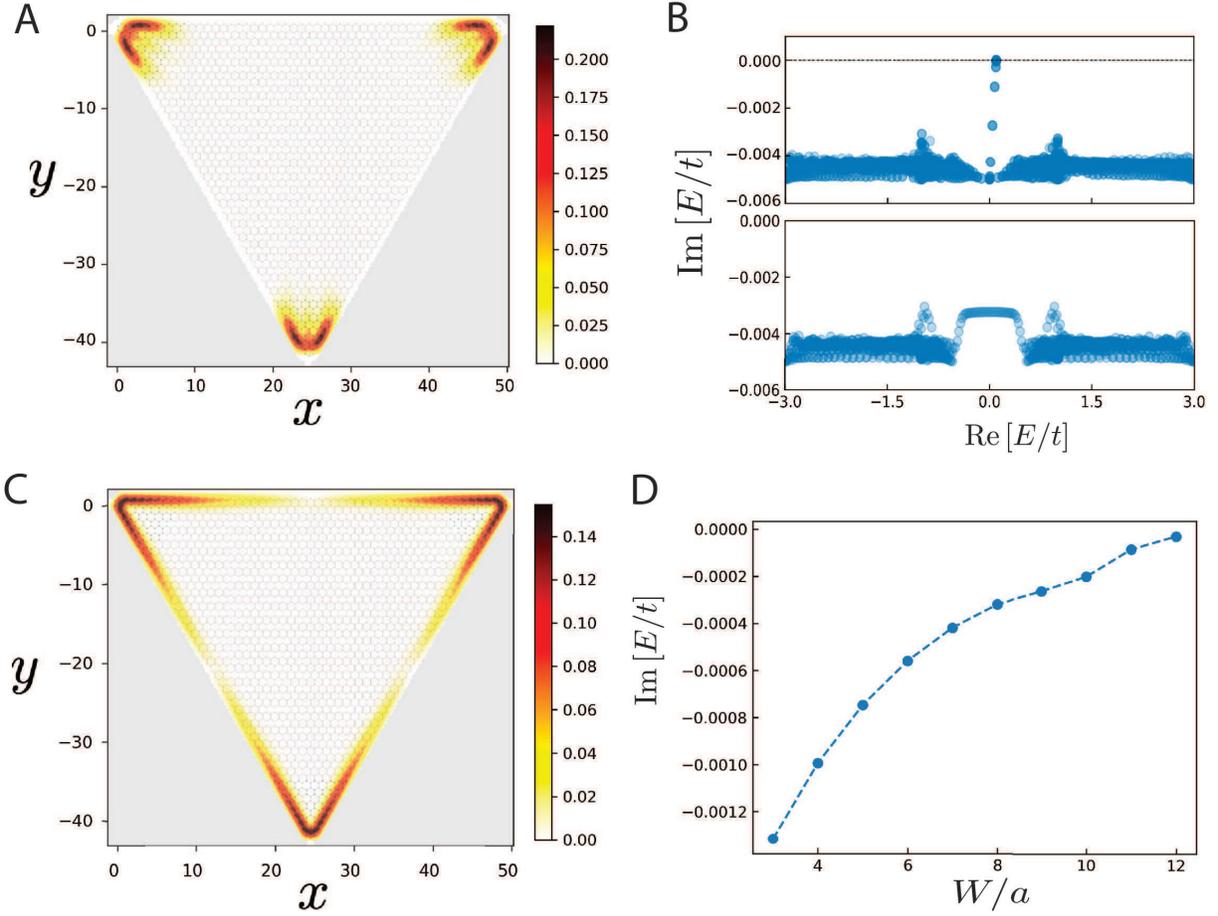}
\caption{ (A) Density of wavefunction for the topological corner state of
 $\pi/6$ zigzag-zigzag corner with a unsymmetric onsite potential
 0.1. (B) Density of wavefunction for the topological corner state of
 $\pi/6$ zigzag-zigzag corner with a next-nearest hopping $0.1t$. (C)
 Complex energy spectra of $\pi/6$ zigzag-zigzag corner samples with a
 unsymmetric onsite potential $0.1t$ (upper panel) and finite imaginary
 next-nearest-neighbor (lower panel). (D) Dependence of imaginary part
 of eigenenergy of the topological corner state in the $\pi/6$
 zigzag-zigzag corner with a real next-nearest hopping 0.1 on the size
 of the non-hermitian mask.} 
\end{center}
\end{figure}

\end{document}